\documentstyle[12pt]{article}
\begin{document}
\begin{titlepage}
\hfill{UM-P-92/52}

\hfill{OZ-92/19}

\hfill{SHEP 91/92}

\hfill{June, 1992}

\vspace{1 cm}
\begin{center}
{\LARGE {\bf Electric Charge Quantization}}\\
\vspace{18 mm}
R. Foot$^{(a)}$\footnote{
Address from Sep.1 - Dec.1, Physical Research Laboratory,
Navrangpura, Ahmedabad-380 009, India.}, H. Lew$^{(b)}$\footnote{
Email: lew@purdd.hepnet, lew\%purdd.hepnet@Csa3.LBL.Gov}
and R. R. Volkas$^{(c)}$\footnote{Email:
U6409503@ucsvc.ucs.unimelb.edu.au}\\
\vspace{5 mm}
\end{center}
{\it (a) Department of Physics,
University of Southampton,
Southampton SO9 5NH,
United Kingdom}
\vspace{5 mm}

\noindent
{\it (b) Physics Department, Purdue University,
West Lafayette IN 47907-1396, U.S.A.}
\vspace{2 mm}

\noindent
{\it (c) Research Centre for High Energy Physics, School of Physics,
University of Melbourne, Parkville 3052, Australia}

\vspace{2 cm}
\centerline{ABSTRACT}

Experimentally it has been known for a long time that
the electric charges of the observed particles appear
to be quantized. An approach to understanding electric
charge quantization that can be used for gauge theories with explicit
$U(1)$ factors -- such as the standard model and its variants --
is pedagogically reviewed and discussed in this article. This approach
uses the allowed invariances of the Lagrangian and their associated
anomaly cancellation equations. We demonstrate
that charge may be de-quantized in
the three-generation standard model with massless neutrinos, because
differences in family-lepton--numbers are anomaly-free. We also
review the relevant experimental limits. Our approach to charge quantization
suggests that the minimal standard model should be extended so that
family-lepton--number differences are explicitly broken. We briefly
discuss some candidate extensions (e.g. the minimal standard model
augmented by Majorana right-handed neutrinos).

\vspace{1.5 cm}
\centerline{\large \it (To appear as a Topical Review in J. Phys.\ G)}

\end{titlepage}
\vskip 2cm

\leftline{\bf 1. Introduction}
\vskip .5cm

The electric charges of all the known elementary particles are integer
multiples (to within experimental errors) of the d-quark charge.
Thus, experimentally, it seems that electric charge is quantized.
A theoretical understanding of electric charge quantization is
important if we hope to understand the interactions of elementary
particles. Of course, understanding elementary
particle interactions is one of the major goals of theoretical physics.

It is not known for certain why electric charge is quantized.
There have been many suggestions over the years, including
higher dimensions [1], magnetic mono-poles [2] and grand unified
theories [3]. All of these suggestions are highly speculative
and very difficult to test experimentally.
Recently, a less speculative approach to electric charge
quantization has emerged. The purpose of
this article to review this approach to understanding the
phenomenon of electric charge quantization [4].

\vskip 1cm
\leftline{\bf 2. The Standard Model}
\vskip .5cm

The starting point is the minimal standard model (MSM), as defined by
its Lagrangian, which is the synthesis of
much work over the last few decades. It appears that the
three non-gravitational forces -- the strong, weak and electromagnetic
-- can be successfully described by a Yang-Mills theory
with the gauge group $G_{SM}$ where
$$G_{SM} = SU(3)_c \otimes SU(2)_L \otimes U(1)_Y. \eqno (1)$$
Under this gauge group, the quarks and leptons
of each generation  transform as
$$Q_L \sim (3, 2, 1/3),\  u_R \sim (3, 1, 4/3),\  d_R \sim (3, 1, -2/3)$$
$$f_L \sim (1, 2, -1), \  e_R \sim (1, 1, -2).  \eqno (2)$$
There is also a Higgs doublet $\phi$ which can be defined
through the Yukawa Lagrangian ${\cal L}_{Yuk}$ where
$${\cal L}_{Yuk} = \lambda_1 \bar f_L \phi e_R + \lambda_2 \bar Q_L \phi d_R
+ \lambda_3 \bar Q_L \phi^c u_R + {\rm H.c.}  \eqno (3)$$
Note that there is an implicit summation over fermion generations in
this equation.
The spin-0 multiplet $\phi$ transforms under $G_{SM}$ as
$$\phi \sim (1, 2, 1), \eqno (4)$$
and $\phi^c \equiv i\tau_2\phi^{*}$.

The Higgs doublet $\phi$ is assumed to be responsible for the fermion
and gauge boson masses. This is achieved by having the Higgs
doublet gain a nonzero vacuum expectation value (VEV), thus
spontaneously breaking $SU(3)_c \otimes SU(2)_L \otimes U(1)_Y$
to $SU(3)_c \otimes U(1)_Q$.
The gauge group $U(1)_Q$ describes the electromagnetic force.
The generator, $Q$, of this Lie group is
the linear combination of $Y$ and the generators $I_i (i=1,2,3)$
of $SU(2)_L$ which is not
broken by the VEV of $\phi$. By the $SU(2)_L \otimes U(1)_Y$ symmetry
of the Lagrangian, we are free to work in the basis in which
the VEV of $\phi$, $\langle \phi \rangle$ has the form
$$\langle \phi \rangle =
\left(\begin{array}{c} 0\\ u
\end{array}\right) \eqno (5)$$
and $u$ is a real number.
Denoting the hypercharge (i.e. the $Y$ charge) of the Higgs doublet as
$y_{\phi}$, we see that the linear combination
$$ Q \equiv c\Bigl(I_3 + {Y \over 2y_{\phi}}\Bigr) \eqno (6)$$
annihilates the VEV given in Eq.(5), for all values of the constant
$c$. Thus, in order to understand the electric charges values
of the quarks and leptons we need to understand each of the $U(1)_Y$
assignments of the quarks and leptons [as can be seen from Eq.(2), there
are five of these values to understand per generation] and also the
value for the $U(1)_Y$ charge of the Higgs doublet, since its value
determines which linear combination of $I_3$ and $Y$
is the electric charge generator.

The overall normalization constant $c$ in Eq.(6)
has no independent physical meaning in the standard model.
Also, the overall normalization of $Y$ has no independent physical
meaning. This is because of a rescaling degree of freedom.
A $U(1)_Y$ gauge theory is invariant under the rescaling
$$ Y \rightarrow \eta Y, \  \  g \rightarrow g/\eta.  \eqno (7)$$
The value of $\eta$ is not physically observable, so we are always free
to change the overall normalization of $Y$ in Eqs.(2) and (4). For
simplicity, we will use this degree of freedom up by
fixing the Higgs boson hypercharge $y_{\phi}$ to be $+1$
in the forthcoming discussion [note that this value for
$y_{\phi}$ obeys convention as per Eq.(4)].
Similarly, we are free to choose the normalization constant $c$ for the
electric charge generator $Q$. It is conventional to set $c$ equal to
one. The result of this is that we are
free to choose the electric charge generator $Q$ to be given by
$$Q = I_3 + Y/2 \eqno (8)$$
without loss of generality. It is important to clearly understand the
origin of Eq.(8). The primary consideration that went into the above
derivation is that the electromagnetic group is defined to be the $U(1)$
invariance left exact after electroweak symmetry breaking. This told us
the key fact that $Q$ was some linear combination of $I_3$ and $Y$.
Having realised this simple but crucial point, we then made the
(trivial) observation that we could conventionally choose
this linear combination to be as given in Eq.(8) by normalizing the
$U(1)$ generators appropriately. The moral of the story is that the
contemptably familiar relation $Q = I_3 + Y/2$ has semantic value; it is
{\it not} just convention.

So, after all of this, we find that
there are five unknowns per generation to be
determined: the quark and lepton hypercharges. We will denote these
through the transformation laws,
$$Q_{mL} \sim (3, 2, y_{m1}), \  u_{mR} \sim (3, 1, y_{m2}), \
d_{mR} \sim (3, 1, y_{m3}),$$
$$f_{mL} \sim (1, 2, y_{m4}), \  e_{mR} \sim (1, 1, y_{m5}), \eqno (9)$$
where $m = 1,...,n_G$ is the generation index. Observe that we do not
intend to make any {\it a priori} assumptions about whether
corresponding multiplets in different generations have the same weak
hypercharges or not. If our goal is to {\it understand} weak
hypercharges, then we should make as {\it few {\em a priori}
assumptions as possible about them}.

Now, we stated at the outset that our approach to electric-charge
quantization would begin with the study of the MSM Lagrangian. Clearly
the generation dependence of the hypercharges in Eq.(9) will be severely
constrained by the requirement of family mixing in the Yukawa
Lagrangian. However, for pedagogical reasons it is best to analyse this
issue one step at a time. We therefore now specialise to the unrealistic
case of just a single quark and lepton generation.

\vskip 1 cm
\leftline{\bf 3. Toy Model \#1}
\vskip 5 mm

Consider the MSM Lagrangian truncated to include only a single fermion
family. This defines our toy model \#1.
Our task is to see how many of the hypercharges $y_{1,2,3,4,5}$
can be determined (we will omit the redundant
generation index for this one generation Toy Model).
There are constraints on these hypercharge values which come
from requiring the Lagrangian to be invariant under the gauge symmetry [5]
(by this we mean in particular, that the Lagrangian must have a $U(1)$
invariance since the gauge group is assumed to have a $U(1)$ factor).
We will refer to these as {\it classical constraints}.
The only part of the Lagrangian which restricts the possible values
of the hypercharges is the Yukawa Lagrangian in Eq.(3).
These Yukawa terms imply the conditions
$$y_1 = y_3 + 1,\  y_1 = y_2 - 1, \ y_4 = y_5 + 1 . \eqno (10)$$
Thus out of the five original unknown hypercharges, only two are left
undetermined by the classical constraints. We will take these as
$y_1$ and $y_4$.

In the case of the MSM, this is as far as the Lagrangian can take us.
However, because we are dealing with a quantum gauge field theory, we
can also use the quantum consistency requirement of {\it gauge anomaly
cancellation} [6]. This requirement
may be justified by either demanding that gauge invariance not
be broken by quantum effects, or by requiring that the standard proof
of renormalizability hold.
Gauge anomalies arise from fermionic
triangle diagrams with gauge bosons
on the external lines. Their amplitudes are proportional to
$$ {\rm Tr} [T^a \{T^b, T^c\}] \equiv A d^{abc}, \eqno (11)$$
where $A$ is a representation-dependent anomaly-coefficient, and
$d^{abc}$ is a set of numbers characteristic of the group.
In this equation, $T^a$, $T^b$ and $T^c$ denote
the generators in the appropriate representations of the Lie
algebra of the gauge group.
Theories are anomalous if the anomaly coefficient
does not vanish when it is summed over the chiral fermions
of the theory (the left- and right-handed fermions enter with
a relative minus sign).

There are two anomaly equations in the MSM
which are independent of
the classical constraints. The first of these arises when
two of the external lines in the triangle graph
are from $SU(2)_L$
gauge bosons with the third being $U(1)_Y$ [7].
The second triangle anomaly arises when all three external
lines are $U(1)_Y$ gauge bosons. These two types of
triangle anomalies are denoted respectively as
$[SU(2)_L]^2 U(1)_Y$ and $[U(1)_Y]^3$.
Evaluating these anomalies by using Eqs.(9-11) leads
to the equations
$$ y_1 = - {y_4 \over 3}\quad {\rm and}\quad
y_4 = -1 \eqno (12)$$
from $[SU(2)_L]^2 U(1)_Y$ and $[U(1)_Y]^3$ anomaly cancellation
respectively.
Thus for the Toy Model case of the MSM restricted to
only one generation we see that
the consistency conditions of gauge invariance of the Yukawa Lagrangian
and anomaly cancellation uniquely imply that electric charge is
quantized. Not only is it quantized, but the quarks and leptons
have the correct electric charges.

\vskip 1 cm
\leftline{\bf 4. Toy Model \#2}
\vskip 5 mm

To facilitate a greater understanding of {\it why} electric
charge is quantized in this one generation model, we consider
the addition of a gauge singlet fermion (the putative
right-handed neutrino) to the above one generation
model:
$$\nu_R \sim (1, 1, y_6). \eqno (13)$$
The Yukawa Lagrangian of Eq.(3) is assumed to contain
the additional term
$$\Delta {\cal L}_{yuk}
= \lambda_4 \bar f_L \phi^c \nu_R + {\rm H.c.} \eqno (14)$$
By following the same steps as in the previous case we might think
that electric charge will still be quantized, since we have one extra
equation [arising from the classical gauge invariance of Eq.(14)]
for the one extra parameter $y_6$.
However, it turns out that the $[U(1)_Y]^3$ gauge anomaly now does not
give an independent constraint. Thus in this case we have only
five equations for six unknowns, and the hypercharges [and hence electric
charges through Eq.(8)] depend on a free continuous parameter,
$\epsilon$, in the following way [5]:
$$Q_L \sim \left(3, 2, {1 \over 3} - {\epsilon\over 3}\right), \
u_R \sim \left(3, 1, {4 \over 3} - {\epsilon\over 3}\right),\
d_R \sim \left(3, 1, - {2 \over 3} - {\epsilon\over 3}\right),$$
$$f_L \sim (1, 2, -1 + \epsilon), \  e_R \sim (1, 1, -2 + \epsilon),\
\nu_R \sim (1, 1, \epsilon). \eqno (15)$$

There is a very easy way to understand the origin of the {\it charge
dequantization} evident in Eq.(15). If one studies this equation,
then it is clear that the
non-standard hypercharge (i.e. the $\epsilon$ bit)
is proportional to $B-L$, where $B$ and $L$ are the usual baryon and lepton
number {\it global} $U(1)$ symmetries [8].
The combination $B-L$ is an anomaly free global symmetry which
is gaugeable. By ``anomaly free'', we mean that $[U(1)_{B-L}]^3,
G_{SM}[U(1)_{B-L}]^2$ and $G_{SM}^2 U(1)_{B-L}$ anomalies
cancel [9]. The symmetry $U(1)_{B-L}$ is also independent of the symmetries
$G_{SM}$. Thus, {\it a priori}, there is nothing to stop
us from gauging any combination
$$ Y \equiv Y_{SM} + \epsilon (B - L) \eqno (16)$$
instead of just $Y_{SM}$ (where $Y_{SM}$ is the standard hypercharge
generator).  In the example of Toy Model \#1 above, where the standard model
is restricted to one generation, the Lagrangian
contains only {\it one} gaugeable $U(1)$ symmetry and
this is precisely $U(1)_{Y_{SM}}$. Thus the consistency
of that theory requires hypercharge (and hence electric charge)
to be quantized to their standard values.

We have thus come to quite a general principle:
\vskip 2mm

\noindent {\it If a Lagrangian
contains global symmetries which are anomaly-free (and hence
gaugeable) and independent of the standard hypercharge $Y$, then
that Lagrangian does not yield electric charge quantization} [5, 10, 11].

\vskip 1 cm
\leftline{\bf 5. Minimal Standard Model with Three Generations}
\vskip 5 mm

Having discussed the toy model case of the standard model
restricted to one generation, we move on to the more interesting
case of the realistic three generation minimal standard model [12, 4].
{}From the principle annunciated above, it is clear that to analyse charge
quantization in this theory we have to find all of its anomaly-free
global $U(1)$ symmetries.

The minimal standard model Lagrangian with three generations has
four global $U(1)$ symmetries. These
are generated by electron-lepton number ($L_e$), muon-lepton
number ($L_{\mu}$), tau-lepton number ($L_{\tau}$) and
baryon number ($B$).
To work out which, if any, combinations of these global symmetries
are anomaly free, we start by considering the most general linear
combination $L'$ where
$$L' \equiv \alpha L_e + \beta L_{\mu}
+ \gamma L_{\tau} + \delta B. \eqno (17)$$
The anomaly constraints are
$$[U(1)_{L'}]^3 \Rightarrow \alpha^3 + \beta^3  + \gamma^3 = 0   $$
$$[SU(2)_L]^2 U(1)_{L'} \Rightarrow
3\delta  + \alpha + \beta + \gamma = 0. \eqno (18)$$
One can check that all other gauge anomaly equations are
not independent of the
above equations. Note that we have only two equations for the four
parameters $\alpha$, $\beta$, $\gamma$ and $\delta$.
So we see that there are an infinite number
of gaugeable global $U(1)$ symmetries in the
standard model. This infinite class can be parameterised as
$$L' = \alpha L_e + \beta L_{\mu} + (- \alpha^3 - \beta^3)^{1/3}L_{\tau} +
{[- \alpha - \beta - (- \alpha^3 - \beta^3)^{1/3}]\over 3} B
\eqno (19)$$
This leads to hypercharge (and hence electric charge) dequantization
through the non-standard formula,
$$Y = Y_{SM} + L'\quad \Rightarrow \quad Q = Q_{SM} + L'/2,  \eqno (20)$$
where $L'$ depends on two continuous free parameters ($\alpha$ and
$\beta$), and $Q_{SM} \equiv I_3 + Y_{SM}/2$ is standard electric
charge.
Equation (20) is the analogue in the three generation MSM of
the Toy Model \#2 result given by Eq.(16). Note that an $\epsilon$
parameter in Eq.(20) (multiplying $L'$)
would be redundant given our definition of $L'$ here.

If one believes in quantum gravity, then one may also wish to impose
the requirement that the mixed gauge-gravitational
anomaly cancel [13, 14].
This requirement is equivalent to
$$Tr\ L' = 0 \eqno (21)$$
which gives the constraint
$$\alpha + \beta + \gamma = 0. \eqno (22)$$
This equation together with Eq.(18) implies that $\delta = 0$, and
either $\alpha$ or $\beta$ or $\gamma$ is equal to zero.
In the first case,
$$L' = L_{\mu} - L_{\tau}, \eqno (23)$$
while
the second case corresponds to
$$L' = L_e - L_{\tau}, \eqno (24)$$
and the last case
corresponds to $$L' = L_e - L_{\mu}. \eqno (25)$$
So, with the extra imposition that the mixed gauge-gravitational anomaly
cancel, we see that
electric charge may be dequantized through [12, 4]
$$Q = I_3 + {Y_{SM} + \epsilon L' \over 2} = Q_{SM} + \epsilon L'/2,
 \eqno (26)$$
where $L' = L_i-L_j, (i,j = e,\mu,\tau; i \neq j)$
[as given in Eqs.(23-25)] and $\epsilon$ is
an arbitrary parameter.

So we conclude that the three generation minimal standard model is not a
straightforward generalization of the one generation Toy Model \#1 as
regards electric charge quantization. Electric charge may be dequantized
in the three generation case but not in the one generation case.
Furthermore, this dequantization can only occur in the ways specified by
Eq.(26) [or Eq.(20) if the mixed gauge gravitational anomaly cancellation
is not assumed].

\vskip 1cm
\leftline{\bf 6. Extensions to the Standard Model}
\vskip .5cm

So we have seen that the minimal standard model with three
generations does not have electric charge quantization. Neither
does the one generation case when a gauge singlet, $\nu_R$, is
added to the fermion spectrum. To obtain electric charge quantization,
one can modify the Lagrangian so that the anomaly free global symmetries
inherent in these models are absent.
For example, the Lagrangian of the one generation model with a gauge
singlet neutrino can be modified by adding the Majorana mass term
${\cal L}_{Maj}$ where
$${\cal L}_{Maj} = M \bar \nu_R (\nu_R)^c. \eqno (27)$$
Since this term breaks the gaugeable global symmetry $U(1)_{B-L}$
[and it doesn't break $U(1)_{Y_{SM}}$] the resulting model will
necessarily have electric charge quantized correctly [10]. Note that the
three generation standard model with Dirac neutrinos has charge
dequantization via $B-L$ only, just like the one generation Toy Model,
because the family-lepton number symmetries are in general explicitly
broken. The three generation case can also be extended by giving
Majorana masses to the right-handed neutrinos in order to ensure charge
quantization.  The above extension of the three generation minimal standard
model (featuring three generations of Majorana right-handed
neutrinos) is our first example of new physics which is specifically
motivated by our approach to electric charge quantization.

In physics there are often many solutions to a given problem,
and the charge quantization problem is no exception.
In the following we will mention a few other
ways in which the three generation
minimal standard model can be extended to achieve electric
charge quantization.

The simplest and most obvious way to modify the three generation
minimal standard model is to add {\it one} right-handed neutrino
singlet with or without a Majorana mass term.
One can show explicitly [15, 16] by
writing down the charged current interactions between the
charged leptons and neutrinos that the individual lepton
numbers--$L_{e, \mu, \tau}$--are no longer in general conserved.
This then leaves
standard hypercharge as the only remaining anomaly free $U(1)$ symmetry
in the model and hence electric charge quantization results.
There are many other extensions of the lepton and neutrino
sector in particular which lead to electric charge quantization [17].

Instead of extending the fermion sector of the minimal standard
model, we can extend the Higgs sector in a way that breaks
those unwanted
gaugeable global symmetries. The simplest extension
to the Higgs sector which can do this is to add another
Higgs doublet (the two-Higgs doublet
model). In this case [16], one finds that the diagonalization of the
charged lepton mass matrix no longer simultaneously diagonalizes
the corresponding Higgs-fermion interactions. In fact there
will now be Higgs-induced flavour changing neutral processes (FCNPs).
[It is usual to introduce some discrete symmetries if one wants the
physical Higgs particles to be less than about a TeV [18] otherwise
these FCNPs will be too large to be compatible with experiment.
However,
from a charge quantization point of view it is desirable that these
discrete symmetries {\it not} be imposed,
since we require generation mixing interactions to break
the gaugeable global symmetries in Eqs.(23-25).]

Another solution, which does not involve mass terms and which
is relevant to
the case where charge dequantization ensues through gaugeable $B-L$,
can be obtained by enlarging the gauge group so that
$U(1)_{B-L}$ is no longer anomaly free [19].

The examples above illustrate our contention that {\it the
methodology we have presented for the understanding of
electric charge quantization serves as a concrete heuristic
guide to model building}. Indeed, we have quoted some specific
extensions of the MSM (e.g. the addition of Majorana right-handed
neutrinos) that can be motivated by our charge quantization
analysis. Of course it should be emphasised that
the only unique, model-independent prediction that one can make from
this electric charge quantization argument is that the global
symmetries $L_e - L_{\mu}$, $L_{\mu} - L_{\tau}$ and $L_e - L_{\tau}$
must be broken.

Furthermore, it is important to note that the method encourages a
``bottom-up'' approach to model building, whereby the successful
low-energy theory (the three generation MSM) is altered in small and
experimentally testable ways. This is to be contrasted
with the traditional approaches to the problem (GUTs, monopoles and
higher dimensions) which introduce quite speculative pieces of new
physics at high and experimentally inaccessible energies.

The critical reader may at this stage comment that our approach is not
in principle completely falsifiable either,
because the parameters defining the
global symmetry breaking terms can be arbitrarily small and still serve
their purpose. For instance, the Majorana mass in Eq.(27) can always be
made small enough to evade experimental bounds. Since
charge quantization is exact no matter how small the Majorana mass is,
we seem to have an unfalsifiable principle.
However, small symmetry breaking
parameters just swap one problem with another. If the terms are absent
and charge is dequantized, the smallness of the parameter $\epsilon$
is a mystery. If the terms are present but the coefficients small,
{\it their} smallness is the mystery. Thus our philosophy successfully
answers this important criticism; the symmetry breaking parameters must
have ``natural'' values.

Before we conclude this section we would like to mention some other
issues. Firstly, the result that the standard model restricted to one
generation (Toy Model \#1) has electric charge quantized correctly is
useful for constructing alternatives to the standard model.
If the standard model emerges as an effective theory at some
scale, then provided all of the non-standard fermions have
$SU(3)_c \otimes SU(2)_L \otimes U(1)_Y$ invariant masses,
charge quantization for the standard fermions
must follow (provided we restrict ourselves to one generation).
This is because fermions with masses that are invariant under
$SU(3)_c \otimes SU(2)_L \otimes U(1)_Y$ do not contribute to the
$SU(3)_c \otimes SU(2)_L \otimes U(1)_Y$ anomaly equations, so that
the Toy Model \#1 result that electric
charge is quantized correctly will
hold. Examples where this result has been used can be found
in Ref.[20].

Also, in all the examples discussed so far we have assumed that symmetry
breaking is due to elementary scalar bosons developing nonzero VEVs.
It may be that some other symmetry breaking scenario occurs in nature.
For example, technicolour is one such possibility. One can of
course analyse models of this kind using the classical constraints
and anomaly conditions as was done for models with elementary scalars [21].

\vskip 1 cm
\leftline{\bf 7. Experimental Constraints on Charge Dequantization}
\vskip 5 mm

In the preceeding section we reviewed some small extensions to the
standard model which ensure electric charge quantization. It is now up to
experiment to discover whether or not any of these extensions
is actually realised in nature. An alternative possibility is that
electric charge is {\it de}quantized. If this is so, then we
know that the parameter $\epsilon$ is constrained to be very small. In
this section we will discuss the various constraints one can derive on
$\epsilon$.

Let us first consider the standard model with massive Dirac neutrinos.
As the discussion above concerning Toy Model \#2 demonstrated, charge
dequantization ensues in this case via
$$Q = Q_{SM} + \epsilon (B-L)/2.  \eqno (28)$$
Note that this is true for any number of generations. This result implies
that both quarks and leptons have non-standard electric charges.

Since the $B-L$ value of the hydrogen atom is zero, electrical
neutrality for hydrogen is maintained even in the face of this form of
charge dequantization. However, neutrons are no longer electrically
neutral, so atoms in general will now be charged. Another unusual
feature of this scenario is charged neutrinos [5].
Experiments on the
neutrality of neutrons [22] yield the stringent bound [5]
$$|\epsilon| < 10^{-21}.  \eqno (29)$$
Such tiny values for $\epsilon$ raise a serious theoretical issue if
charge is in fact dequantized in nature. Why should nature choose such a
small value, when the theoretical consistency of the theory allows any
value? This type of naturalness puzzle suggests that it is more likely
for nature to have chosen an extension of the standard model which
guarantees charge quantization, rather than opting for dequantization.
However, we should be circumspect in treating theoretical prejudices
such as this as inviolate principles; only experiment can provide the
ultimate answer.

What are the bounds on charge dequantization in the three generation
minimal standard model [23]? Recall that in this case the electric charge
formula is given by Eq.(26), and therefore only leptons have
non-standard charges (given the adopted normalization for $Q$
and the aforementioned insistence on mixed gauge-gravitational
anomaly cancellation). Also,
charge dequantization is generation-dependent, with one lepton family
retaining canonical charges. For this model, the experimental constraints
depend critically on whether the first generation leptons are chosen to
have standard charges (the $L_{\mu} - L_{\tau}$ case) or not (the $L_e -
L_{\mu,\tau}$ cases).

Consider first dequantization via either $L_e - L_{\mu}$ or $L_e -
L_{\tau}$. The hydrogen atom, and ordinary matter in general,
now acquires a nonzero charge. Direct experimental measurements on
atomic neutrality yield the bounds [23, 24]
$$|\epsilon| < 10^{-17} - 10^{-21}. \eqno (30)$$
More stringent, but less rigorous, bounds may be derived by considering
the effects of a charged planet earth. The most severe example comes
from requiring that the radial electric field near the earth's surface
be less than about 100 V/m, yielding $|\epsilon| < 10^{-27}$ [23, 25].
Unfortunately this bound assumes that the number of protons in the earth
equals the number of electrons, so it cannot be taken as rigorous. (One
can also derive bounds on dequantization via $B-L$ from the effects of a
charged earth [5]. The bounds in this case are more rigorous because the
neutron is charged, and so an overall charge for the earth does not
arise from a delicate cancellation between proton and electron charges.)

The $L_{\mu} - L_{\tau}$ case is more interesting, because the bounds
are many orders of magnitude smaller.
Ordinary matter is neutral in this model
because first generation quarks and
leptons have their standard charges. The first bound we will quote comes
from determining the maximum allowed charge difference between the
electron and the muon. This is derived by considering the 1-loop photonic
contribution to the anomalous magnetic moments of these two particles.
Due to the celebrated precision of the agreement between the standard
theoretical predictions for these quantities and their measured values,
the constraint $|\epsilon| < 10^{-6}$ must be imposed in order
to preserve this success. A better bound is obtained by considering the
effects of charged muon-neutrinos in $\nu_{\mu}$-$e$ scattering
experiments. By requiring the new photon exchange contribution to this
process to be smaller than the experimental uncertainty on the
cross-section, the bound
$$|\epsilon| < 10^{-9}  \eqno (31)$$
can be derived [23]. As far as we are aware,
this is the best constraint one can derive from terrestrial experiments.

If we allow ourselves to also consider astrophysical and cosmological
phenomena, somewhat more stringent bounds can be obtained. Since
$\nu_{\mu}$ and $\nu_{\tau}$ are charged and massless, massive
plasmon states in red giant stars can decay into charged
neutrino-antineutrino pairs,
which subsequently escape the star and so carry
off energy. By requiring that the rate of energy loss per unit volume
due to these unorthodox processes not exceed the nuclear energy
generation rate per unit volume, the bound $|\epsilon| < 10^{-14}$ is
obtained [23, 26]. A cosmological constraint is obtained by noting that photons
will acquire a nonzero thermal electric mass from interactions with the
charged relic neutrino background plasma [23].
The induced electric mass for
photons will result in an effective long-distance violation of Gauss'
Law for classical electric fields. The best experimental test of Gauss'
Law [27] yields the bound $|\epsilon| < 10^{-12}$.
Both of the bounds quoted
in this paragraph rely on the correctness of standard astrophysics and
cosmology for their veracity. Of course, since it is impossible to test
astrophysical and cosmological models in nearly as much detail as the
standard model of particle physics, we can never be sure that we really
understand stellar objects and the universe properly. We would
therefore have much more confidence in the rigor of the bounds quoted
in the previous paragraph than those in the present paragraph.

\vskip 1 cm
\leftline{\bf 8. Concluding Remarks}
\vskip 5 mm

Before concluding this review, we would like to mention some other work
that has been motivated by electric charge quantization: The two
simplest $Z'$ models one can construct are immediately evident if one
understands our approach to charge quantization. If no right-handed
neutrinos exist, then $Z'$ bosons coupling to family-lepton number
differences can exist. The detailed phenomenology of these models has
recently been considered in the literature [28]. If, on the other hand,
neutrinos have Dirac masses then a $Z'$ boson coupling to $B-L$ can be
introduced [29]. Both of these $Z'$ models are simple, because no exotic
fermions need to be introduced in order to cancel the gauge anomalies (unless
you consider right-handed neutrinos to be exotic).

The discussion of charge dequantization has motivated
some work on charge nonconservation [25, 30].
Charge dequantization may occur as a result of charge
nonconservation. Usually the constraints on the photon mass are so
tight as to rule out the possibility of observing charge dequantization if
charge non-conservation is the only source of charge dequantization in
the model. However, it has been shown [30] that there are
models which have observable charge dequantization arising from electric
charge nonconservation.

In conclusion then, we have reviewed recent work on a simple,
falsifiable, bottom-up approach to electric charge quantization. We
have demonstrated how the consistency of the three generation minimal
standard model fails to ensure charge quantization.
Several small extensions of the miminal standard model constructed to
deliver exact charge quantization were then discussed, and constraints
on the unorthodox prospect of charge dequantization were presented.
\vskip 2cm
\centerline{\bf Acknowledgements}

RRV is supported by the Australian Research Council through a Queen
Elizabeth II Fellowship. HL is supported by the United States DOE.

\newpage
\vskip 1cm
{\LARGE \bf REFERENCES}
\vskip 1cm
\noindent
[1] O. Klein, Nature {\bf 118}, 516 (1926).

\vskip .5cm
\noindent
[2] P. A. M. Dirac, Proc.\ Roy.\ Soc.\ A{\bf 133}, 60 (1931).

\vskip .5cm
\noindent
[3] J. C. Pati and A. Salam, Phys.\ Rev.\ D{\bf 10}, 275 (1974);
H. Georgi and S. L. Glashow, Phys.\ Rev.\ Lett. {\bf 32}, 438 (1974).
Also of relevance is: L. B. Okun, M. B. Voloshin and V. I. Zakharov,
Phys.\ Lett.\ B{\bf 138}, 115 (1984).

\vskip 5mm
\noindent
[4] For an earlier review see
R. Foot, G. C. Joshi, H. Lew and R. R. Volkas, Mod.\ Phys.\ Lett.\
A{\bf 5}, 2721 (1990). The present review discusses
more recent developments. See the earlier review for a comparison
between the approach to charge quantization presented here and some
similar, but less successful, methodologies.

\vskip 5mm
\noindent
[5] R. Foot, G. C. Joshi, H. Lew and R. R. Volkas, Mod.\ Phys.\ Lett.\ A{\bf
5}, 95 (1990); (E) {\it ibid.} 2085 (1990).
Similar observations were made independently and much earlier by
N. G. Deshpande, Oregon preprint OITS-107 (1979) (unpublished).

\vskip .5cm
\noindent
[6] S. L. Adler, Phys.\ Rev.\ {\bf 177}, 2426 (1969); J. S. Bell and
R. Jackiw, Nuovo\ Cimento\ A{\bf 60}, 49 (1969); S. L. Adler and W. Bardeen,
Phys.\ Rev.\ {\bf 182}, 1517 (1969);
C. Bouchiat, J. Illiopoulos and P. Meyer, Phys.\ Lett.\ B{\bf 38}
519 (1972); D. Gross and R. Jackiw, Phys.\ Rev.\ D{\bf 6}, 447 (1972);
H. Georgi and S. L. Glashow, Phys.\ Rev.\ D{\bf 6}, 429 (1972).

\vskip .5cm
\noindent
[7] Note that the anomaly cancellation equations are the same whether or
not spontaneous gauge symmetry breaking occurs. This allows us to work
in the weak eigenstate basis here.

\vskip 5mm
\noindent
[8] Note that we use the conventional normalization for
the global symmetries $L$ and $B$. This means that leptons have
$L=1$ while quarks and other particles have zero lepton number. Also,
quarks have $B=1/3$ while
the leptons and other particles have zero baryon number.

\vskip .5cm
\noindent
[9] Note that for a $U(1)$ {\it gauge} symmetry to be anomaly-free, we
must require that the $[U(1)]^3$ anomaly cancel. A confusion in this
regard can arise from the fact that an anomaly-free
{\it continuity equation} for
the symmetry current of a {\it global} $U(1)$ symmetry can be obtained
even if the $[U(1)]^3$ anomaly is nonzero. The term
``anomaly-free'' is used in two different senses
in the above two sentences.

\vskip .5cm
\noindent
[10] K. S. Babu and R. N. Mohapatra, Phys.\ Rev.\ Lett. {\bf 63}, 938 (1989);
Phys.\ Rev.\ D{\bf 41}, 271 (1990).

\vskip .5cm
\noindent
[11] R. Foot, Int.\ J.\ Mod.\ Phys. A{\bf 6}, 1467 (1991).

\vskip 5mm
\noindent
[12] R. Foot, Mod.\ Phys.\ Lett.\ A{\bf 6}, 527 (1991).

\vskip 5mm
\noindent
[13] R. Delbourgo and A. Salam, Phys.\ Lett.\ B{\bf 40}, 381 (1972);
T. Eguchi and P. Freund, Phys.\ Rev.\ Lett.\ {\bf 37}, 1251 (1976);
L. Alvarez-Gaum\'e and E. Witten, Nucl.\ Phys.\ B{\bf 234}, 269 (1983).

\vskip 5mm
\noindent
[14] Note that the mixed gauge-gravitational anomalies in Toy models
\#1 and \#2 are not independent of the classical and gauge
anomaly cancellation constraints.

\vskip 5mm
\noindent
[15] R. Foot and S. F. King, Phys.\ Lett.\ B{\bf 259}, 464 (1991).

\vskip .5cm
\noindent
[16] J. Sladkowski and M. Zralek, Phys.\ Rev.\ D{\bf 45}, 1701 (1992).

\vskip .5cm
\noindent
[17] See for example, D. C. Choudhury and U. Sarkar, Phys.\ Lett.\ B{\bf 268},
96 (1991).

\vskip .5cm
\noindent
[18] S. L. Glashow and S. Weinberg, Phys.\ Rev.\ D{\bf 15}, 1958 (1977);
E. A. Paschos, Phys.\ Rev.\ D{\bf 15}, 1966 (1977).

\vskip 5mm
\noindent
[19] R. Foot, Madison preprint MAD/TH/89-6 (unpublished).

\vskip 5mm
\noindent
[20] R. Foot, Phys.\ Rev.\ D{\bf 40}, 3136 (1989);
R. Foot and O. F. Hern\'andez, Phys.\ Rev.\ D{\bf 41}, 2283 (1990);
R. Foot and H. Lew, Phys.\ Rev.\ D{\bf 41}, 3502 (1990);
Nuovo\ Cimento,\ A{\bf 104}, 167 (1991); R. Foot, Mod.\ Phys.\
Lett.\ A{\bf 5}, 1947 (1990).

\vskip .5cm
\noindent
[21] X.-G. He, G. C. Joshi, H. Lew, B. H. J. McKellar and
R. R. Volkas, Phys.\ Rev.\ D{\bf 40}, 3140 (1989); R. Foot, H. Lew
and R. R. Volkas, Phys.\ Rev.\ D{\bf 42}, 1851 (1990).

\vskip 5mm
\noindent
[22] J. Baumann, J. Kalus, R. Gahler and W. Mampe, Phys.\ Rev.\ D{\bf
37}, 3107 (1988).

\vskip 5mm
\noindent
[23] K. S. Babu and R. R. Volkas, Melbourne preprint UM-P-92/27, Bartol
Research Institute preprint BA-36 (Phys.\ Rev.\ D, in press);
see also E. Takasugi and M. Tanaka,
Prog.\ Theor.\ Phys.\ {\bf 87}, 679 (1992).

\vskip 5 mm
\noindent
[24] M. Marinelli and G. Morpugo, Phys.\ Lett.\ B{\bf 137}, 439 (1984).

\vskip 5mm
\noindent
[25] K. S. Babu and R. N. Mohapatra, Phys.\ Rev.\ D{\bf 42}, 3866
(1990).

\vskip 5mm
\noindent
[26] S. Davidson, B. Campbell and D. Bailey, Phys.\ Rev.\ D{\bf 43},
2314 (1991).

\vskip 5mm
\noindent
[27] E. R. Williams, J. E. Faller and H. A. Hill, Phys.\ Rev.\ Lett.\
{\bf 26}, 721 (1971).

\vskip 5mm
\noindent
[28] X.-G. He, G. C. Joshi, H. Lew and R. R. Volkas, Phys.\ Rev.\ D{\bf
43}, R22 (1991); {\it ibid.} D{\bf 44}, 2118 (1991).

\vskip 5mm
\noindent
[29] For recent phenomenological analyses of gauged $B-L$,
see X.-G. He, G. C. Joshi and R. R. Volkas, Phys.\ Lett.\ B{\bf 240},
441 (1990); S. L. Glashow and U. Sarid, Phys.\ Rev.\ Lett.\ {\bf 64},
725 (1990); Phys.\ Rev.\ D{\bf 42}, 3224 (1990).

\vskip 5mm
\noindent
[30] E. Takasugi and M. Tanaka, Phys.\ Rev.\ D{\bf 44}, 3706 (1991); M.
Maruno, E. Takasugi and M. Tanaka, Prog.\ Theor.\ Phys.\ {\bf 86}, 907
(1991);  M. Tanaka, Doctoral thesis, Osaka preprint OS-GE 16-91.

\end{document}